\documentclass{ws-procs9x6}

\def\beq{\begin{equation}}
\def\eeq{\end{equation}}
\def\bea{\begin{eqnarray}}
\def\eea{\end{eqnarray}}
\def\ben{\begin{enumerate}}
\def\een{\end{enumerate}}

\def\a{\alpha}

\def\d{\delta}

\def\phi{\varphi}

\def\s{\sigma}
\def\t{\tau}

\def\gsim{\; \raisebox{-.8ex}{$\stackrel{\textstyle >}{\sim}$}\;}

\def\half{{\textstyle{\frac{1}{2}}}}

\begin{document}

\title{Einstein-\AE ther Gravity:\\
Theory and Observational Constraints}

\author{Ted Jacobson}

\address{Center for Fundamental Physics\\
University of Maryland \\
College Park, MD 20742-4111, USA\\
E-mail: jacobson@umd.edu}

\maketitle

\abstracts{Einstein-\ae ther theory is general relativity
coupled to a dynamical unit timelike vector field.
A brief review of current theoretical
understanding and observational constraints on the
four coupling parameters of the theory is given.}

\section{Introduction}

In general relativity (GR), spacetime structure is
determined by a dynamical metric tensor field
$g_{ab}$ and nothing else, and the theory is
both diffeomorphism invariant and locally
Lorentz invariant. Einstein-\ae ther theory is the
extension of GR that incorporates
a dynamical unit timelike vector field
$u^a$---the ``\ae ther"---which breaks the local
Lorentz symmetry down to a 3d rotation subgroup.
Direct coupling of matter to the \ae ther would
violate local Lorentz symmetry yet preserve
diffeomorphism invariance.
This paper presents a brief overview of the
current theoretical and observational status of this theory,
assuming that matter does not couple directly to the
\ae ther.

The action involving metric and \ae ther is
highly constrained. Besides the cosmological constant
term, the only independent diffeomorphism invariant
local  terms containing no more than two derivatives are
\beq S = -\frac{1}{16\pi G}\int \sqrt{-g}~ (R+K^{ab}_{mn}
\nabla_a u^m \nabla_b u^n)~d^{4}x, \label{action} \eeq
where
$R$ is
the Ricci scalar, $K^{ab}_{mn}$ is defined as
\beq K^{ab}_{mn} = c_1 g^{ab}g_{mn}+c_2\d_m^a\d_n^b
+c_3\d_n^a\d_m^b +c_4u^au^bg_{mn} \eeq
with dimensionless coupling constants $c_i$,
and the unit timelike constraint on the \ae ther is
implicit.
(The metric signature is $({+}{-}{-}{-})$
and the speed of light defined by the metric $g_{ab}$
is unity.)
Higher derivatives would be suppressed by powers of
a (presumably) small length, e.g. the
Planck length. 
It is assumed here that the
\ae ther is aligned at large scales with the rest frame of
the microwave background radiation.

Einstein-\ae ther theory---``\ae -theory" for short---is
similar to the vector-tensor
gravity theories studied by Will and
Nordvedt,\cite{willnord} but with the crucial
difference that the vector field is constrained to
have unit norm. This constraint eliminates a
wrong-sign kinetic term for the length-stretching
mode,\cite{Elliott:2005va} hence gives the theory a
chance to be viable. An equivalent theory using the
tetrad formalism was first studied by
Gasperini,\cite{Gasperini} and in the above form it was
introduced by Jacobson and Mattingly.\cite{Jacobson:2000xp}

\section{Newtonian and post-Newtonian limits}
\label{NPN}

In the weak-field, slow-motion limit \ae -theory reduces to
Newtonian gravity,\cite{Carroll:2004ai} with a value of  Newton's
constant $G_{\rm N}$ related to the parameter $G$ in the action
(\ref{action})  by
\beq G_{\rm N}=\frac{G}{1-c_{14}/2}, \label{GN}\eeq
where $c_{14}\equiv c_1+c_4$. (Similar notation
is  used below for other additive combinations
of the $c_i$.)
For any choice of
the $c_i$, all parameterized post-Newtonian
(PPN) parameters\cite{willLR} of \ae -theory
agree with those of
GR\cite{Eling:2003rd,Foster:2005dk}
except the preferred frame
parameters $\alpha_{1,2}$
which are given by\cite{Foster:2005dk}
\bea
    \alpha_1&=& \frac{-8(c_3^2 + c_1c_4)}{2c_1 - c_1^2+c_3^2}\label{alpha1}\\
    \alpha_2&=&\frac{\a_1}{2}
    -\frac{(c_1+2c_3-c_4)(2c_1+3c_2+c_3+c_4)}{c_{123}(2-c_{14})}\label{alpha2} \eea
(This particular way of expressing $\a_2$ was given in
Ref.\ \refcite{Foster:2006az}. The small $c_i$ form of $\a_2$ was first
computed in Ref.\ \refcite{Graesser:2005bg}.)

Observations currently impose the strong constraints
$\a_1 \lesssim 10^{-4}$ and $\a_2\lesssim 4\times
10^{-7}$.\cite{willLR} Since \ae -theory has four free
parameters $c_i$, we may set
$\alpha_{1,2}$ exactly
zero by imposing the conditions\cite{Foster:2005dk}
\begin{eqnarray}
 c_2&=&(-2c_1^2-c_1c_3 + c_3^2)/3c_1 \label{zeroalphac2}\\
 c_4&=&-c_3^2/c_1 \label{zeroalphac4}.
\end{eqnarray}
 With
(\ref{zeroalphac2},\ref{zeroalphac4}) satisfied, {\it
all} the PPN parameters of \ae -theory are equivalent to
those of GR. (The parameters $\a_{1,2}$ can also be
set to zero by imposing $c_{13}=c_{14}=0$, but this
case is pathological, as discussed in section
\ref{special}.)

\section{Homogeneous isotropic cosmology}

Assuming spatial homogeneity and isotropy, $u^a$
necessarily coincides with the 4-velocity of the
isotropic observers,
and the \ae ther stress tensor
is just a certain combination of
the Einstein tensor and the stress tensor of a
perfect fluid with energy density proportional to the
inverse square of the scale factor, like the curvature
term in the Friedman equation.\cite{Mattingly:2001yd,Carroll:2004ai}
The latter contribution plays no important cosmological
role since the spatial curvature is small, while the former
renormalizes the gravitational constant appearing in
the Friedman equation, yielding\cite{Carroll:2004ai}
\beq
G_{\rm cosmo}=\frac{G}{1+(c_{13}+3c_2)/2}.
\eeq
Since $G_{\rm cosmo}$
is not the same as $G_{\rm N}$ the expansion rate
of the universe differs from what would have been expected
in GR with the same matter content. The ratio is constrained
by the observed primordial ${}^4$He abundance to satisfy
$|G_{\rm cosmo}/G_{\rm N} - 1|<1/8$.\cite{Carroll:2004ai}
When the PPN parameters
$\a_{1,2}$ are set to zero by (\ref{zeroalphac2},\ref{zeroalphac4}),
it turns out that $G_{\rm cosmo}=G_{\rm N}$,
so this nucleosynthesis constraint is automatically satisfied.\cite{Foster:2005dk}

\section{Linearized wave modes}

When linearized about a flat metric and constant
\ae ther, \ae -theory posesses five massless modes
for each wave vector: two spin-2,
two spin-1, and one spin-0 mode.
The squared speeds of these modes
relative to the \ae ther rest frame
are
given by\cite{Jacobson:2004ts}
\bea\label{speeds}
\mbox{spin-2}\qquad&&1/(1-c_{13})\label{s2}\\
\mbox{spin-1}\qquad&&(c_1-\half c_1^2+\half
c_3^2)/c_{14}(1-c_{13})\label{s1}\\
\mbox{spin-0}\qquad&&c_{123}(2-c_{14})/c_{14}(1-c_{13})(2+c_{13}+3c_2)\label{s0}
\eea
The corresponding polarization tensors
were found in one gauge in Ref.\ \refcite{Jacobson:2004ts} and
in another gauge in Ref.\ \refcite{Foster:2006az}. 
The energy density of
the spin-2 modes is always positive, while for the
spin-1 modes it has the sign of $(2c_1 -c_1^2 +c_3^2)
/(1-c_{13})$, and for the spin-0 modes it has the sign
of $c_{14}(2-c_{14})$.\cite{Eling:2005zq,Foster:2006az}
(These reduce to the results of Ref.\ \refcite{Lim:2004js}
in the decoupling limit where
gravity is turned off.) 

These squared speeds correspond to
(frequency/wavenumber)${}^2$,
so must be non-negative to avoid
imaginary frequency instabilities.
They must moreover be greater than unity
(super-luminal),
to avoid the existence of vacuum \v{C}erenkov
radiation by matter.\cite{Elliott:2005va}
(The strongest constraints arise from
the existence of ultra high energy
cosmic rays.) And the mode energy
densities should be positive, to avoid dynamical
instabilities.
With the $\a_{1,2}=0$ conditions
(\ref{zeroalphac2},\ref{zeroalphac4}) imposed,
all of these conditions are met for all of the modes
if and only if
$c_\pm=c_1\pm c_3$ are restricted
by the inequalities\cite{Foster:2005dk}
\bea\label{superluminal}
0&\le& c_+\le1\label{sl1}\\
0&\le& c_-\le c_+/3(1-c_+).\label{sl2}
\eea
Interestingly, if the mode speeds are
instead required to be {\it less}
than unity (sub-luminal), then
the spin-1 and spin-0 energy
densities are negative. Hence not only the
\v{C}erenkov constraint, but
also energy positivity (together with
$\a_{1,2}=0$)
requires mode speeds greater than unity.

Note that when (\ref{zeroalphac4}) holds, we have
$c_{14}=2c_+c_-/(c_++c-)$, which satisfies $0\le
c_{14}<2$ when the constraints (\ref{sl1},\ref{sl2})
hold. Thus in particular the condition for attractive
gravity mentioned in section \ref{NPN} need not be
separately imposed, and $c_{14}$ is non-negative.

\section{Primordial perturbations}

Given the same
$G_{\rm N}$,  and assuming the PPN parameters
$\a_{1,2}$ vanish, the primordial power in cosmological
spin-0 and spin-1 perturbations
is unchanged relative to GR, while
the power in spin-2 perturbations
differs from that in GR by the factor
$(1-c_{14}/2)(1-c_{13})^{1/2}$.\cite{Lim:2004js,Li:2007vz}
When the
constraints (\ref{sl1},\ref{sl2}) are satisfied this
factor is smaller than unity,
hence these spin-2 perturbations are
even more difficult to detect than in
GR.
As for the late time evolution
of these perturbations,
neutrino stresses in the radiation dominated epoch
source the spin-1 mode, which leads to modified
matter and CMB spectra. The effect is rather small however,
and is degenerate with matter-galaxy bias and with
neutrino masses.\cite{Li:2007vz}

\section{Radiation damping and strong self-field effects}

If the fields are weak everywhere
(including inside the radiating bodies),
 and
the PPN parameters $\a_{1,2}$ vanish,
radiation is sourced only by the quadrupole.
Waves of spins 0, 1 and 2 are radiated, 
and the
net power
is given by
$(G_N {\mathcal A}/5)\dddot{Q}_{ij}^2$, where
$Q_{ij}$ is the quadrupole moment and
 ${\mathcal A}={\mathcal A}[c_i]$ is a function
 of the coupling parameters $c_i$
that reduces to unity in the case of GR.\cite{Foster:2006az}
Agreement with
the damping rate of
GR (confirmed to $\sim 0.1\%$ in
binary pulsar systems\cite{willLR})
can be achieved by imposing the condition
${\mathcal A}[c_i]=1$,
which is consistent with the
constraints (\ref{sl1},\ref{sl2}).

Compact sources with strong internal fields such
as neutron stars or black holes
can be handled\cite{Foster:2007gr} using an
``effective source" dynamics specified by a
worldline action
integral
\beq
S=-m_0\int d\t\; [1 +\s(v^au_a-1)+\s'(v^au_a-1)^2+\dots],
\label{effectiveaction}
\eeq
where $v^a$ is the 4-velocity of the body, $u_a$ is
the local background value of the \ae ther, and $\s$
and $\s'$  are
constants characterizing the body, called a
``sensitivity parameters" or just ``sensitivities".
The sensitivites scale as $c_i$ for small $c_i$.

The effects of nonzero sensitivities on two-body
dynamics and radiation rates lead to a number of
phenomena that are constrained by observations,
including violations of the strong equivalence
principle, modifications of the post-Newtonian
dynamics, modifications of quadrupole sourced
radiation, and both monopole and dipole sourced
radiation. When $\a_{1,2}=0$, all of these constraints
are met provided the sensitivities are less than
$\sim 0.001$, which will certainly be the case if
$c_i\lesssim 0.01$.\cite{Foster:2007gr}\footnote{This
corrects an error in
version 1 of  Ref.\ \refcite{Foster:2007gr}, where
$\s$ is said to scale as $c_i^2$. (Also the
a prefactor $c_{14}$ in Eqn. (70) should be deleted.)
As a result of this correction,
the likely constraints on $c_i$ are an order
of magnitude stronger, as stated here.\cite{bzf-pc}}
To be more
precise would require knowing the actual dependence of the
sensitivities on the $c_i$, which has so far only been
determined for $\s$ and only at leading order (where 
$\s$ vanishes
when $\a_{1,2}=0$).
(The speed $V$ of the
observed binaries with respect to the background
\ae ther frame can be neglected in formulating these
constraints provided
$V\lesssim 10^{-2}$, which is easily satisfied for
any known proper motion relative to the rest frame
of the microwave background radiation.\cite{Foster:2007gr})

\section{Spherically symmetric stars and black holes}

Unlike GR, \ae -theory has a spherically symmetric mode,
corresponding to radial tilting of the \ae ther. For each mass,
there is a two parameter family of spherically
symmetric static vacuum solutions, rather than
a unique solution as in GR.\cite{Eling:2006df}
Asymptotic flatness reduces this to a one parameter
family.\cite{Eling:2003rd,Eling:2006df}
The solution outside a static
star is the unique solution
for a given mass in which
the \ae ther is aligned with the Killing vector.\cite{Eling:2006df}
This ``static \ae ther"
vacuum solution
depends
on the $c_i$ only through the combination
$c_{14}$, and
was found analytically (up to
inversion of a transcendental equation).\cite{Eling:2006df}
It is stable to linear
perturbations under the same conditions as for
stability of flat
spacetime, with the exception of the case
$c_{123}=0$.\cite{Seifert:2007fr}

The solution inside a fluid star has been found by
numerical integration, both for constant
density\cite{Eling:2006df} and for realistic neutron
star equations of state.\cite{Eling:2007xh}
The maximum masses
for neutron stars range from about 6 to 15\% smaller
than in GR when $c_{14}=1$,
depending on the equation of state.
The corresponding surface redshifts can be
as much as 10\% larger than in GR for the same mass.
Measurements of high
gravitational masses or precise surface redshifts thus
have the potential to yield strong joint constraints
on $c_{14}$ and the equation of state.
The radius of the innermost stable circular orbit (ISCO)
differs from the GR value  $6G_{\rm N}M$
by a small term of relative order about $0.03c_{14}$.

For black holes,
the condition of regularity at the spin-0 horizon
selects a unique solution from the one-parameter
family for a given mass.\cite{Eling:2006ec}
When a black hole forms from collapse of matter, the
spin-0 horizon develops in a nonsingular region of
spacetime, where the evolution should be regular. This
motivated the conjecture that collapse
produces a black hole
with nonsingular spin-0 horizon, which has been
confirmed for some particular examples
in numerical simulations of collapse of
a scalar field.\cite{Garfinkle:2007bk}

The black holes with nonsingular spin-0 horizons are
rather close to Schwarzschild outside the horizon for
a wide range of couplings; for instance, the ISCO
radius differs by a  factor $(1 + 0.043 c_1 +
0.061 c_1^2)$, in
the case with $c_3=c_4=0$ and $c_2$ fixed so that the
spin-0 speed is unity.\cite{cte-pc} (This 
expansion is accurate at least when $c_1\le0.5$. 
No solution with
regular spin-0 horizon exists in this case when $c_1
\gsim 0.8$.) Inside the horizon the solutions differ
more, but like Schwarzschild they contain a spacelike
singularity. Black hole solutions with singular spin-0
horizons have been studied in Ref.\
\refcite{Tamaki:2007kz}. These solutions can differ
much more outside the horizon. Quasi-normal modes of
black holes in \ae -theory have been investigated in
Refs.\ \refcite{Konoplya:2006rv}.

\section{Special values of $c_i$?}
\label{special}

The first case to be examined in
detail\cite{Kostelecky:1989jw,Jacobson:2000xp} was
$c_{13}=c_2=c_4=0$, i.e.\ the ``Maxwell action"
together with the unit constraint on the vector. The
PPN result for $\a_2$ (\ref{alpha2}) is infinite in
this case, and the spin-0 mode speed is zero. The
perturbation series used in the PPN analysis is thus
evidently not applicable. Independently of that
however, other problems with this case have been
identified, such as the formation of shock
discontinuities\cite{Jacobson:2000xp,Clayton:2001vy}
and a possibly related
instability.\cite{Seifert:2007fr}

Assuming now that
$\a_{1,2}=0$ and the
constraints (\ref{sl1},\ref{sl2}) are satisfied,
and putting aside the case $c_1=c_3=0$ which
is not covered by existing PPN analyses, all
but one of the
cases in which one of the $c_i$ vanishes, or
in which one of $c_{13}$, $c_{14}$, or $c_{123}$
vanishes, have the property that the spin-1 mode speed
(\ref{s1}) diverges while the energy of that mode is
nonzero. It seems very unlikely that such cases are
observationally viable, although they have not been examined
carefully. The exception is the special case
$c_3=c_4=2c_1+3c_2=0$, with $2/3<c_1<1$. This large
value of $c_1$ is probably inconsistent with the
strong field constraints from orbital binaries, but as
mentioned above those are not yet precisely known
because the sensitivity parameters have not yet been
computed for neutron stars, so this case is not yet
ruled out.

\section{Conclusion}

Einstein-\ae ther theory is an intriguing theoretical laboratory
in which gravitational effects of possible
Lorentz violation can be meaningfully studied.
There is a large (order unity) two-parameter space of Einstein-\ae ther
theories for which (i) the PPN parameters are identical to those of GR,
(ii) the linear perturbations are stable and carry positive energy,
(iii) there is no vacuum \v{C}erenkov radiation,
(iv) the dynamics of the cosmological scale factor and
perturbations differ little from GR,
(v) non-rotating neutron star and black hole solutions are
close to those of GR, but might be distinguishable
with future observations. Radiation damping from binaries,
imposes an order $0.001$ constraint on one combination
of the parameters.
Strong self-field effects in neutron stars and black holes
produce violations of the strong equivalence principle and
higher order post-Newtonian effects which
will constrain all the
parameters $c_i$ to be less than
around $0.01$, presuming that
the sensitivity parameters for neutron stars (which have not
yet been computed with the required precision) turn out to
have the expected magnitude.

\section*{Acknowledgments}
I am grateful to C.T.\ Eling,
B.Z.\ Foster, B.\ Li, and E.\ Lim for helpful correspondence.
This work was
supported by NSF grant PHY-0601800.

\end{document}